\documentstyle{article}

\newcommand{\order}{{\cal O}}
\newcommand{\pv}{{\bf p}}
\newcommand{\be}{\begin{equation}}
\newcommand{\ee}{\end{equation}}
\newcommand{\bea}{\begin{eqnarray}}
\newcommand{\eea}{\end{eqnarray}}
\newcommand{\nl}{\nonumber \\}
\newcommand{\action}{{\cal S}}

\newcommand{\psib}{\overline{\psi}}
\newcommand{\chib}{\overline{\chi}}

\title{Flavor-Symmetry Restoration and Symanzik Improvement for Staggered
Quarks}

\author{G.~Peter~Lepage\\
\small Newman Laboratory of Nuclear Studies\\
\small Cornell University, Ithaca, NY 14853-5001}

\date{\small September 1998}

\begin{document}

\maketitle

\begin{abstract}
We resolve contradictions in the literature concerning the origins and
size of unphysical flavor-changing strong interactions generated by the
staggered-quark discretization of QCD. We show that the leading
contributions are tree-level in~$\order(a^2)$ and that they can be
removed by adding three correction terms to the link operator in the
standard action. These corrections are part of  the systematic
Symanzik improvement of the staggered-quark action. We present a
new improved action for staggered quarks that is accurate
up to errors of~$\order(a^4,a^2\alpha_s)$\,---\,more accurate than
most, if not all, other discretizations of light-quark dynamics.
\end{abstract}

\section{Introduction}

The staggered-quark discretization of QCD has several advantages for
numerical simulations. In particular, numerical evaluation of quark
propagators is far more efficient with staggered quarks than with any
other standard discretization. This is especially true for small quark
masses where a chiral symmetry of the staggered-quark action prohibits
additive mass renormalizations, in close analogy with continuum QCD.
This chiral symmetry also implies that errors caused by the nonzero
grid spacing,~$a$, are automatically quadratic in~$a$, rather than
linear as in Wilson's formulation. These advantages are offset by the
fact that staggered quarks always come in groups of four identical
flavors, which is too many for QCD. The multiplication of flavors is
not in itself a major problem; the effective number of flavors is
easily adjusted in simulations. The real problem is that the standard
staggered-quark action allows relatively large flavor-changing strong
interactions, which are completely absent in real~QCD. These
flavor-changing interactions greatly complicate the interpretation of
staggered-quark simulations; a deeper understanding of their origins is
essential for precision work with staggered quarks, particularly at
larger lattice spacings.

Recently, in~\cite{gpl-tsukuba,sinclair}, it was observed that the
dominant flavor-changing interaction is due to one-gluon exchange
between quarks. It was also observed that this interaction is a
short-distance lattice artifact of relative order~$a^2$ which can be
removed by tree-level modifications of the lattice action. A more
recent study, however, contradicts this conclusion by arguing that the
one-gluon mechanism is redundant in~$\order(a^2)$, and therefore that
flavor-changing interactions are higher order in~$a$
or~$\alpha_s$\,\cite{luo-onshell}.  In this paper we resolve this
contradiction by showing that one-gluon exchange is indeed responsible
for the dominant flavor-changing interaction, as originally argued;
the flavor-changing effects are not redundant in~$\order(a^2)$.
Furthermore, these effects can be canceled by a simple modification
of the quark-gluon interaction, similar to but more complicated than
the ``fat-link'' improvement shown to reduce flavor-changing effects
in~\cite{milc-fatlink}. Our fat link is similar to that
in~\cite{sinclair}, but less complicated.

Fat-link improvement was first introduced as a heuristic scheme for
reducing flavor-changing interactions in staggered quarks.  In fact it
is part of the systematic Symanzik improvement of the
lattice quark action. As mentioned above, the standard action has no
errors in~$\order(a)$ (for light quarks). 
There are only two sources of~$\order(a^2)$ error
at tree-level. One is the flavor-changing interaction due to one-gluon
exchange; the other is a kinetic term common to all standard
quark discretizations\,\cite{naik,luo-onshell}. In this paper, we show
how to correct for both of these. The result is a lattice quark action
that is accurate up to errors of~$\order(a^4,a^2\alpha_s)$\,---\,the
most accurate to date\,\cite{d234}.

\section{Flavor-Changing Interactions}

The staggered-quark formalism is derived from 
the simplest discretization of the quark action in QCD, the ``naive''
quark action:
\be
\action = \sum_{x}\, \psib(x)\,(\gamma\cdot\Delta +m)\,\psi(x).
\ee
Here $\Delta_\mu$ is a discrete version of the covariant derivative,
\be \label{der}
\Delta_\mu\,\psi(x) \equiv
\frac{1}{2au_0}\left(U_\mu(x)\,\psi(x+a\hat{\mu})
-U^{\dagger}_\mu(x-a\hat{\mu})\,\psi(x-a\hat{\mu})\right),
\ee
$U_\mu(x)$ is the gluon link-field, and $u_0$ is the mean-link tadpole
improvement\,\cite{gpl-ti,gpl-tsukuba}. 
This action has an exact ``doubling'' symmetry under the
transformation
\bea
\psi(x)\quad \to \quad \tilde{\psi}(x) &\equiv& i\gamma_5\gamma_\rho\,
(-1)^{x_\rho/a}\,\psi(x) \nl
&=& i\gamma_5\gamma_\rho\,\exp(i\,x_\rho\pi/a)\,\psi(x).
\eea
Thus any low energy-momentum mode, $\psi(x)$, of the theory is equivalent to
another mode, $\tilde{\psi}(x)$, that has momentum $p_\rho \approx
\pi/a$, the maximum allowed on the lattice. This new mode is one of the
``doublers'' of  the naive 
quark action. The doubling transformation can be
applied successively in two or more directions; the general
transformation is 
\be
\psi(x)\quad\to\quad\tilde\psi(x)\,\equiv\,
\prod_\rho\,\left(i\gamma_5\gamma_\rho\right)^{\zeta_\rho}\,\,
\exp(i\,x\cdot\zeta\,\pi/a)\,\psi(x)
\ee
where $\zeta$ is a vector with one or more components equal to~1 and
all the others~0. Consequently there are 15~doublers in all (in four
dimensions), which we label with the fifteen different~$\zeta$'s. 

As a consequence of the doubling symmetry, the standard low-energy
mode and the fifteen doubler modes must be interpreted as sixteen
equivalent flavors of quark. (The sixteen flavors are reduced to four
by staggering the quark field, as we discuss at the end.) This unusual
implementation of quark flavors has surprising consequences. Most
striking is that a low-energy quark that absorbs momentum close to
$\zeta\,\pi/a$, for one of the fifteen~$\zeta$'s, is not driven far off
energy-shell. Rather it is turned into a low-energy quark of another
flavor. Thus the simplest process by which a quark changes flavor is
the emission of a single gluon with momentum~$q\approx\zeta\,\pi/a$. 
This gluon is highly virtual, with $q^2= \order((\pi/a)^2)$, and therefore 
it must immediately be reabsorbed by another quark, whose flavor will
also change. Flavor changes necessarily involve highly virtual gluons,
and so are perturbative for typical lattice spacings. Consequently
one-gluon exchange, with gluon momentum $q\approx\zeta\,\pi/a$, is
the dominant flavor-changing interaction since it is lowest order
in~$\alpha_s(\pi/a)$.

Flavor-changing gluon exchanges between quarks are in effect identical to
flavor-changing quark-quark contact interactions since the exchanged 
gluons are highly virtual. Thus the effects of these exchanges
can be canceled by adding 
four-quark interactions to the quark action\,\cite{gpl-tsukuba}. The
contact terms required at tree-level were explicitly constructed
in~\cite{luo-onshell}; these could be added to the staggered-quark
action and the leading flavor-changing interactions removed. There is,
however, a simpler modification of the action that accomplishes the
same goal. This is to change the quark-gluon coupling in the naive
action so as to suppress gluon momenta near $\zeta\,\pi/a$ for each of
the~$\zeta$'s\,\cite{stagg}. 
For example, defining a covariant second derivative that acts on link
operators, 
\bea
\Delta^{(2)}_\rho\,U_\mu(x) &\equiv&
\frac{1}{u_0^2\,a^2} \left(
U_\rho(x)\,U_\mu(x+a\hat\rho)\,U_\rho^\dagger(x+a\hat\mu) 
- 2\,u_0^2\,U_\mu(x) \right. \nl
&&\quad\quad + \left.U_\rho^\dagger(x-a\hat\rho)\,U_\mu(x-a\hat\rho)
\,U_\rho(x-a\hat\rho+a\hat\mu)\right),
\eea
the operator 
\be
\left(1+\sum_{\rho\ne\mu}\frac{a^2\Delta^{(2)}_\rho}{4}\right)\,U_\mu(x)
\ee
is identical to the link operator for low gluon momenta, up to errors
of~$\order(a^2)$, but vanishes when a single gluon with momentum
$q_\rho = \pi/a$ is extracted. Replacing the links in the naive
action by this operator would remove flavor-changing interactions with
$\zeta^2=1$ (that is, one component equal to~1 and all others~0). 
This is essentially what
was done in the fat-link improvement scheme presented
in~\cite{milc-fatlink}. To remove all leading-order flavor-changing
interactions, we replace
\be
U_\mu(x) \quad\to\quad V_\mu(x) \equiv 
\left(1+\sum_\zeta (1-\zeta_\mu)\,c(\zeta^2)\,{\cal P}(\zeta)\right)\,U_\mu(x)
\ee
in the naive action, where 
\be \label{pdef}
{\cal P}(\zeta) \equiv \left.{\prod_\rho}
\left(\frac{a^2\,\Delta^{(2)}_\rho}{4}\right)^{\zeta_\rho}
\,\right|_{\rm symmetrized}
\ee
is symmetrized over all possible orderings of the operators.
We drop corrections to~$V_\mu$ with $\zeta_\mu=1$ since the other parts of
the corresponding 
quark-gluon vertex vanish when the gluon
has $q_\mu=\pi/a$ (just as in the original action).
The coefficients $c(\zeta^2)$ have perturbative expansions:
\be
c(\zeta^2) = 1 + \order(\alpha_s(\pi/a)).
\ee
Tree-level values, with tadpole-improved operators,
should be sufficiently accurate for most
applications at typical lattice
spacings. Alternatively, we can tune the coefficients
nonperturbatively, for example 
to remove flavor splittings between pions. (In four
dimensions there are three independent $c$'s just as there are three
independent pion splittings for
small quark masses~\cite{sharpe-splittings}.) 
Such tuning maximizes the cancellation of residual tadpole effects.
Taking just the tree-level values for the couplings, $V_\mu$ simplifies to
\be
V_\mu(x) = \left.\prod_{\rho\ne\mu}
\left(1+\frac{a^2\,\Delta^{(2)}_\rho}{4}\right)\right|_{\rm symm.}
U_\mu(x)\quad\mbox{(tree-level)}.
\ee
Here the prefactors vanish when acting on a gluon field that has any
momentum component other than $q_\mu$ equal to $\pi/a$; as mentioned
above, momenta with
$q_\mu=\pi/a$ are suppressed by the remainder of the quark-gluon
vertex. The action in~\cite{sinclair} is similar to ours, but has
three additional operators, for suppressing $q_\mu=\pi/a$, that are
unnecessary.

Our prescription for removing the leading flavor-changing
interactions is to replace the link field~$U_\mu$ in the naive action
by the link field~$V_\mu$ defined above. As in the standard case, our
improved naive action is
equivalent to four identical, uncoupled staggered-quark theories.
Three of these can be removed by staggering the quark spinors over
nearby lattice sites, thereby
reducing the number of flavors to four (in four dimensions).
Our final staggered theory is identical to the standard 
theory but with the substitution: $U_\mu\to
V_\mu$. The additional computing required to implement this
improvement is negligible.

\section{$\order(a^2)$ Symanzik Improvement}

As mentioned in the previous section, the leading flavor-changing
interactions can be canceled by adding four-quark contact interactions
to the lattice action. This means that the errors introduced by these
interactions are~$\order(a^2)$ at tree-level in the original
theory\,\cite{gpl-tsukuba,sinclair}. In~\cite{luo-onshell} the
relevant contact interactions were converted into quark bilinears
by transforming the gluon field in the path integral. The
bilinears that resulted were of the form
\be
a^2\,\psib(x)\gamma_\mu \left({\cal P}(\zeta)\Delta^{(2)}U_\mu(x)
\right) \psi(x+a\hat\mu).
\ee
Such operators appear to be of relative order~$a^4$ and higher, and thus
they were dropped in the~$\order(a^2)$ analysis
of~\cite{luo-onshell}. It was argued that the original contact terms
could be transformed away, that they were redundant in~$\order(a^2)$. 
This is incorrect. The quark bilinear, despite
appearances, affects the theory in~$\order(a^2)$. While the factors of
$a^2\Delta^{(2)}_\rho$ acting on the link field~$U_\mu(x)$ in the
bilinear operator strongly suppress
quark-gluon interactions for small gluon momenta, these same factors
are of order unity for the large, flavor-changing gluon momenta that
leave the quark on-shell. Thus the flavor-changing parts of these terms
are of the same order as the flavor-changing part of the original
quark-gluon vertex; that is, they are~$\order(a^2)$.

Conventional power-counting, as assumed in~\cite{luo-onshell} and
elsewhere, is incorrect for a theory in which a highly virtual gluon
can connect two on-shell quarks. Just about any quark bilinear will
contribute flavor-changing interactions of this sort. In such
situations additional factors of $a^2\Delta^{(2)}_\rho$ acting on the
gluon field bring no additional suppression of the bilinear's
contribution.  The correct power counting for the flavor-changing part
of a quark bilinear can always be determined by examining the
corresponding quark-quark contact terms.

The $\order(a^2)$~corrections introduced when $U_\mu$ is replaced by
$V_\mu$ cancel all tree-level flavor-changing interactions in that
order. It is straightforward to remove the remaining errors in that
order to obtain an $\order(a^2)$~Symanzik improved quark
action. First we must cancel a flavor-{\em conserving} order~$a^2$ error
due to the $\zeta^2=1$ parts of~$V_\mu$. This is removed by further
modifying~$V_\mu$:
\be
V_\mu(x)\quad \to \quad V_\mu^\prime(x) \,\equiv\,
V_\mu(x) - \sum_{\rho\ne\mu}\frac{a^2(\Delta_\rho)^2}{4}\, U_\mu(x).
\ee
Here the $(\Delta_\rho)^2$'s 
are squares of discretized covariant first-order
derivatives for links, defined analogously to~Eq.~(\ref{der}); they
cancel the low-energy effects of the single $\Delta_\rho^{(2)}$'s in~$V_\mu$
without affecting their high-momentum behavior. Second we must correct
the discretization of the derivative through~$\order(a^2)$
by replacing
\be
\Delta\cdot\gamma \quad\to\quad  (\Delta_\mu -
\frac{a^2}{6}\,(\Delta_\mu)^3)\,\gamma_\mu
\ee
in the action\,\cite{naik,luo-onshell}. While the first of these terms
requires the $V_\mu^\prime(x)$ link field in place of~$U_\mu(x)$,
this is unnecessary in the $\Delta^3_\mu$~term. The additional first-order
derivatives in that term connect quark fields separated by an even
number of lattice spacings, and therefore they suppress both
flavor-conserving 
and flavor-changing interactions (to $\order(a^4,a^2\alpha_s)$). 
Thus our $\order(a^2)$-accurate action for naive quarks is 
\be
\action = \sum_{x}\, \psib(x)\,\left(\gamma\cdot\Delta^\prime
-\frac{a^2}{6}\gamma\cdot\Delta^{3}+m\right)\,\psi(x)
\ee
where
\be
\Delta_\mu^\prime\,\psi(x) \equiv
\frac{1}{2au_0}\left(V_\mu^\prime(x)\,\psi(x+a\hat{\mu})
-V^{\prime\,\dagger}_\mu(x-a\hat{\mu})\,\psi(x-a\hat{\mu})\right).
\ee
The quark fields for this action can again be staggered in the
standard fashion to obtain an $a^2$-accurate 
staggered-quark action for only four
flavors of quark (in four dimensions):
\be 
\action_{\rm stagg.} = \sum_{x}\,
\eta_\mu(x)\,\chib(x)\left(\Delta^\prime_\mu 
-\frac{a^2}{6}\Delta^{3}_\mu\right)\chi(x) + m \chib(x)\chi(x)
\ee
where $\chi(x)$ is a one-component quark field and 
\be
\eta_\mu(x) \equiv (-1)^{x_1/a+\cdots+x_{\mu-1}/a}.
\ee

We have already noted that the coupling constants in~$V_\mu$ (and
$V_\mu^\prime$) can be
tuned nonperturbatively, if desired, to remove flavor-splittings
between different pions in the staggered theory. The kinetic
correction can also be tuned nonperturbatively using a symmetry, in
this case Lorentz invariance (relativity). For example, the
coefficient of this term might be tuned to adjust
\be
c^2(\pv) \equiv \frac{E^2(\pv)-m^2}{\pv^2}
\ee
to $c^2=1$ for the pion with the lowest nonzero momentum.

\section{Conclusions}
In this paper we have established that the leading flavor-changing
interactions in the naive and staggered-quark actions are
short-distance, tree-level, $\order(a^2)$ effects. They are most
easily removed by modifying the link field in the original action. The
minimal number of correction terms required is three, one for each of
the pion splittings at small quark mass. Other schemes for fattening
the link operators have shown some success at reducing
flavor-splittings\,\cite{milc-fatlink,sinclair}. 
Our scheme should be at least as successful, and probably substantially
more so if the coefficients of the correction terms are tuned to
remove all residual tadpole effects (for example, by 
completely removing the pion splittings).

The flavor-changing interactions in the standard action are not
redundant at tree-level in~$\order(a^2)$. They must be corrected to
obtain an $a^2$-accurate, Symanzik-improved action. The fat link we
introduce for this purpose is not heuristic, but rather a consequence
of the systematic Symanzik improvement of the action. In this paper we
presented a new Symanzik-improved action for staggered quarks that
is accurate through tree-level in~$\order(a^2)$. As such it is more
accurate than most other discretizations of quark dynamics, and is therefore
a prime candidate for use on lattices with large lattice spacings, in
the range 0.1--0.3\,fm.

We thank H. Trottier, S. Sharpe, and P. Mackenzie for useful
conversations. This 
work was supported by a grant from the National Science
Foundation.


\begin{thebibliography}{10}
\bibitem{stagg} Similar schemes to that discussed here 
are proposed in~\cite{sinclair} and implied by the analysis
in~\cite{luo-onshell}. Our
formalism is simpler than either of these. Our final formalism is
almost identical to the last one proposed in~\cite{milc-fatlink},
which appeared while we were preparing this manuscript; we tune the
couplings differently and also remove all tree-level
$\order(a^2)$~errors.
\end{thebibliography}
\end{document}